\documentclass[usenatbib]{mnras}

\usepackage{graphicx}	% Including figure files
\usepackage{subfigure}
\usepackage{amsmath}	% Advanced maths commands
\usepackage{amssymb}	% Extra maths symbols
\usepackage{multicol}        % Multi-column entries in tables
\usepackage{bm}		% Bold maths symbols, including upright Greek
\usepackage{pdflscape}	% Landscape pages

\usepackage[T1]{fontenc}
\usepackage{ae,aecompl}

\usepackage{newtxtext,newtxmath}
\title[Constraints on cosmological models from quasars]{\textit{Constraints on Cosmological Models from Quasars Calibrated with Type Ia Supernova by a Gaussian Process}}

\author[H.  Zhang et al.]{Haixiang Zhang $^{1}$, Yang Liu\thanks{Co-first author} $^1$, Hongwei Yu $^{1,5}$,\thanks{hwyu@hunnu.edu.cn}, Xiaodong Nong $^{2,3,4}$, Nan Liang\thanks{liangn@bnu.edu.cn} $^{2,3,4}$  and Puxun Wu $^{1,5}$\thanks{pxwu@hunnu.edu.cn}
\\
$^{1}$Department of Physics and Synergistic Innovation Center for Quantum Effects and Applications, \\ Hunan Normal University, Changsha, Hunan 410081, China\\
$^{2}$Key Laboratory of Information and Computing Science Guizhou Province, Guizhou Normal University, Guiyang, Guizhou 550025, China\\
$^{3}$School of Cyber Science and Technology, Guizhou Normal University, Guiyang, Guizhou 550025, China\\
$^{4}$Joint Center for FAST Sciences Guizhou Normal University Node, Guiyang, Guizhou 550025, China\\
$^{5}$Institute of Interdisciplinary Studies, Hunan Normal University, Changsha, Hunan 410081, China	
}	

\pubyear{2023}

\begin{document}
\label{firstpage}
\pagerange{\pageref{firstpage}--\pageref{lastpage}}
\maketitle

\begin{abstract}
In this paper, we use  quasars calibrated from  type Ia supernova (SN~Ia) to   constrain  cosmological models.
We consider three different X-ray luminosity ($L_{X}$) - ultraviolet luminosity ($L_{UV}$) relations of quasars, i.e.,   the standard $L_{X}$-$L_{UV}$ relation and  two redshift-evolutionary relations (Type~I and Type~II) respectively constructed from copula and considering a redshift correction to the luminosity of quasars.
Only in the case of the Type I relation,  quasars can always  provide effective constraints on the $\Lambda$CDM model.
Furthermore, we show that, when  the observational Hubble data (OHD) are added,  the constraints on the absolute magnitude $M$ of SN~Ia  and the Hubble constant $H_0$ can be obtained. In the $\Lambda$CDM model, the OHD measurements plus  quasars with the Type I relation yields  $M$ =$-19.321^{+0.085}_{-0.076}$, which is  in good agreement with the measurement from SH0ES ($M=-19.253\pm{0.027}$),
and   $H_0$ = $70.80\pm3.6~\mathrm{km~s^{-1}Mpc^{-1}}$, falling  between the measurements  from SH0ES and the Planck cosmic microwave background radiation data. 
%Additionally,  both the Akaike information criterion  and the Bayesian information criterion  show that  quasars prefer the Type I  relation.
\end{abstract}

\begin{keywords}
Quasars: general - \emph{(cosmology:)} dark energy - cosmology: observations
\end{keywords}

%%%%%%%%%%%%%%%%%%%%%%%%%%%%%%%%%%%%%%%%%%%%%%%%%%

%%%%%%%%%%%%%%%%% BODY OF PAPER %%%%%%%%%%%%%%%%%%

\section{Introduction}

 The $\Lambda$CDM (cosmological constant $\Lambda$ plus  cold dark matter) model   is the simplest cosmological model to explain the accelerating cosmic  expansion.  Although consistent with many observations~\citep{Perlmutter1999,Riess1998,eBOSS2021,Brout2022}, it still suffers the  Hubble tension~\citep{Riess2020,Valentino2021,Perivolaropoulos2022,Dainotti2021,Dainotti2022a,Liu2023}, which refers to the more than $5\sigma$ disagreement between the constraints on the Hubble constant $H_0$ from the nearby type Ia supernova (SN Ia) and the  cosmic microwave background (CMB) radiation data, respectively. Using the latest SN~Ia measurements, \cite{Riess2022} obtain $H_0=73.04\pm1.04~\mathrm{km~s^{-1}Mpc^{-1}}$ in  a model-independent manner, while  the CMB data from the Planck 2018 survey yield $H_0=67.36\pm0.54~\mathrm{km~s^{-1}Mpc^{-1}}$ \citep{Planck2020} in the framework of the $\Lambda$CDM model. Many other observational data have been utilized to search  the possible origins of the $H_0$ tension, but a satisfactory explanation remains elusive.  Since the redshift range of commonly used observations, including SN~Ia~\citep{Scolnic2022},  observational Hubble parameter data (OHD)~\citep{Moresco2020}, baryon acoustic oscillations~\citep{Eisenstein2005}, and strong gravitational lenses~\citep{Suyu2010, Suyu2013}, reaches only about $z\sim2$, while the CMB data is near $z\sim1100$, cosmological data in the mid-redshift region ($2\la z\la 1100$) might offer important insights into the origins of the Hubble tension.

Quasars (quasi-stellar objects) are the extremely luminous and persistent sources of light found in the universe, which can be detected  at  $z>7$~\citep{Mortlock2011,Banados2017,Wang2021}. To use quasars as  the standard candles  in cosmology, the luminosity relation of quasars needs to be constructed.  In this regard,
 an empirical non-linear relation between the ultraviolet (UV) luminosity and the X-ray luminosity $(L_{X}$-$L_{UV})$ has been proposed in \citep{Tananbaum1979, Zamorani1981, Avni1986, Risaliti2015}, and it has been widely applied to investigate the high-redshift universe using quasars~\citep{Risaliti2015,Risaliti2019,Lusso2016,Lusso2017,Lusso2019,Lusso2020,Velten2020,Khadka2020a,Khadka2020b,Khadka2021,Khadka2023,Wei2020,Li2021,Li2023,Hu2022a,Bargiacchi2022,Dainotti2023,Yang2020}.   Recently, \cite{Khadka2020a} showed some evidences of redshift evolution of the X-ray and UV relation. By considering a power-law redshift correction ($(1+z)^k$) to the quasar luminosity, \cite{Dainotti2022} obtained a three-dimensional and redshift-evolutionary version of the $L_{X}$-$L_{UV}$ relation.  In \citep{Dainotti2022}, the coefficient of the redshift correction term is determined by using the Efron-Petrosian (EP) method~\citep{Efron1992}. This coefficient will be treated as a free parameter  in the following analysis. \cite{Wang2022} introduced an improved three-dimensional $L_{X}$-$L_{UV}$ relation with a redshift-dependent term given by a powerful statistical tool called copula\footnote{Copula is a powerful tool developed in modern statistics to describe the correlation between multivariate random variables~\citep{Nelsen2007} and it has been used to construct the  redshift-evolutionary relation of Gamma-ray bursts~\citep{Liu2022a,Liu2022b}.}. The increased reliability of the $L_X$-$L_{UV}$ relations as a cosmological tool has been established  in \citep{Wang2022, Wang2024, Dainotti2022,Lenart2023}
 
Similar to the Gamma-ray burst (GRB) cosmology, the quasar cosmology also suffers the so-called \textit{circularity problem} since a fiducial cosmological model is usually  assumed  to calibrate the empirical relation when quasars are used to constrain the  cosmological models. 
Inspired by the idea of distance ladder with the Cepheids and SN~Ia, the \textit{low-redshift calibration} has been proposed to calibrate the GRB relations from  SN~Ia to build the Hubble diagram of GRB~\citep{Liang2008,Liang2010,Liang2011,Kodama2008,Capozziello2008,Wei2009,Wei2010,Liu2015,Wang2016,Demianski2017,Liu2022b,Liang2022,Li2023,Mu2023,Xie2023}.   In addition, the simultaneous fitting or global fitting method has also been established to avoid the {\it circularity problem} in the GRB cosmology~\citep{Ghirlanda2004,Li2008,Dainotti2023b, Cao2024}. Here, we want to apply the low-redshift calibration method  in the quasar cosmology and calibrate the luminosity relations of quasars by using  SN Ia.\footnote{ 
As demonstrated in \citep{Wang2022, Lenart2023}, it has been established that certain cosmological parameters, such as $\Omega_{\mathrm m0}$, are linearly independent of the correction coefficients. Hence, there is no impediment to constraining these parameters without calibration. However, calibration can significantly improve the precision of cosmological parameter estimation.} To achieve the luminosity distance at the redshift of quasars from  SN Ia, we need to
 employ a cosmological model-independent method to reconstruct the Hubble diagram of SN~Ia.  In this regard, let us note that the Gaussian process (GP) is a fully Bayesian statistical method used for reconstructing the Hubble diagram from existing data and predicting unknown data \citep{Seikel2012a}, which can effectively reduce errors in the reconstructed results.
In recent years, the GP method has found extensive applications in the fields of cosmology and astrophysics \citep{Seikel2012b,Busti2014,Yu2016,Lin2018,Wei2018,Yu2018,Pan2020,Hu2022b}.

Thus, in this paper,  the GP method will be used to reconstruct the Hubble diagram of  the 1590 Pantheon+ SN~Ia data points, which are obtained by removing the data in the redshift region of $z<0.01$ from  1701 SN light curves of 1550 spectroscopically confirmed SN Ia with the maximum redshift being about $ z \sim 2.3$ ~\citep{Scolnic2022}, and then  the luminosity distances of low-redshift quasars, which are the subset of  the dataset comprising 2421 X-ray and UV flux measurements of quasars~\citep{Lusso2020},  can be derived from the SN~Ia Hubble diagram.  From an initial dataset of 21,785 data points, a total of 2,421 quasars have been selected. These quasars originate from seven different samples, observed using instruments such as \textit{Chandra} and \textit{XMM-Newton}. Quasars displaying UV reddening, significant host-galaxy contamination in the near-infrared, or poor photometry data were excluded. Additionally, adjustments for Eddington bias were considered. Following these corrections, \cite{Lusso2020} compiled the final, refined sample of 2,421 quasars, covering a redshift range from 0.009 to 7.52. It is important to note that no  $K$-correction was applied to the quasars \citep{Bloom2001, Lusso2020}. These low-redshift quasars can be used to determine the coefficients in different $L_X$-$L_{UV}$ relations. Extrapolating  the results to the high-redshfit quasars,  we can obtain the luminosity distances of all quasars model-independently, and use these distances to constrain the cosmological models. 
Since the quasar sample only cannot give constraints on the absolute magnitude ($M$) of SN~Ia and the Hubble constant ($H_0$) simultaneously, we also add  32 OHD measurements  obtained by the cosmic chronometers  method \citep{Moresco2020}, which relate to the evolution of differential ages of passive galaxies at different redshifts \citep{Jimenez2002},
to obtain  constraints on $M$ and  $H_0$.

The rest of this paper is organized as follow. In Section \ref{sec2}, we respectively calibrate, by using the low-redshift calibration method, three different $L_{X}$-$L_{UV}$ relations  from SN~Ia:  the standard relation and two redshift-evolutionary relations (Type~I and Type~II) constructed from copula and considering a redshift correction to the luminosity of quasars.
The constraints on different cosmological models from  quasars and OHD are given in  Section \ref{sec3}. Section \ref{sec4} shows the discussion on results.
Conclusions are summarized in Section \ref{sec5}.

\section{Calibrating the $L_{X}$-$L_{UV}$ Relations from SN Ia}\label{sec2}
The standard $L_{X}$-$L_{UV}$ relation, which is a non-linear relation between the X-ray luminosity $(L_{X})$ and the  UV luminosity $(L_{UV})$  of quasars~\citep{Tananbaum1979, Zamorani1981, Avni1986, Risaliti2015}, takes the form
\begin{equation}\label{eq:std_L}
	\log \left(L_{X}\right)=\beta+\gamma \log \left(L_{U V}\right).
\end{equation}
Here  $\beta$ and $\gamma$ are two coefficients , and ``log'' denotes the logarithm to  the base $10$. Expressing the luminosity in terms of the flux, one can obtain
\begin{eqnarray}\label{eq:std_F}
	\log \left(F_{X}\right)&=&2(\gamma-1) \log \left(d_{L}\right)+\beta+(\gamma-1) \log (4 \pi) \nonumber \\
	&&+\gamma \log \left(F_{U V}\right).
\end{eqnarray}
 Here  $F_X=\frac{L_X}{4\pi d_L^2}$ and $F_{UV}=\frac{L_{UV}}{4\pi d_L^2}$ are the observed flux of X-ray and UV, respectively, and $d_L$ is the luminosity distance, which contains the information of cosmological models.

Recently, the standard $L_{X}$-$L_{UV}$ relation has been generalized  to contain possible redshift-evolutionary effects:
\begin{eqnarray}\label{eq:evo_F}
	\log \left(F_{X}\right)&=&2(\gamma-1) \log \left(d_{L}\right)+\beta+(\gamma-1) \log (4 \pi) \nonumber \\
	&& +\gamma \log \left(F_{U V}\right)+\alpha \ln(\bar{\alpha}+z).
\end{eqnarray}
Here $\alpha$ is a coefficient, and $\alpha\neq 0$ represents that the relation evolves with redshift. The constant $\bar{\alpha}$ equals to $1$ or $5$. When  $\bar{\alpha}=5$, the relation given in Eq.~(\ref{eq:evo_F}), which is named  the Type I relation in this paper, corresponds to that constructed by using the copula function~\citep{Wang2022}. 
If $\bar{\alpha}=1$, the corresponding relation is called  the Type II relation, 
and is obtained by assuming  that the luminosity of quasars is corrected  via a redshift-dependent function $(1+z)^k$~\citep{Dainotti2015,Dainotti2022}.

\begin{figure}
	\centering
	\includegraphics[width=0.95\columnwidth]{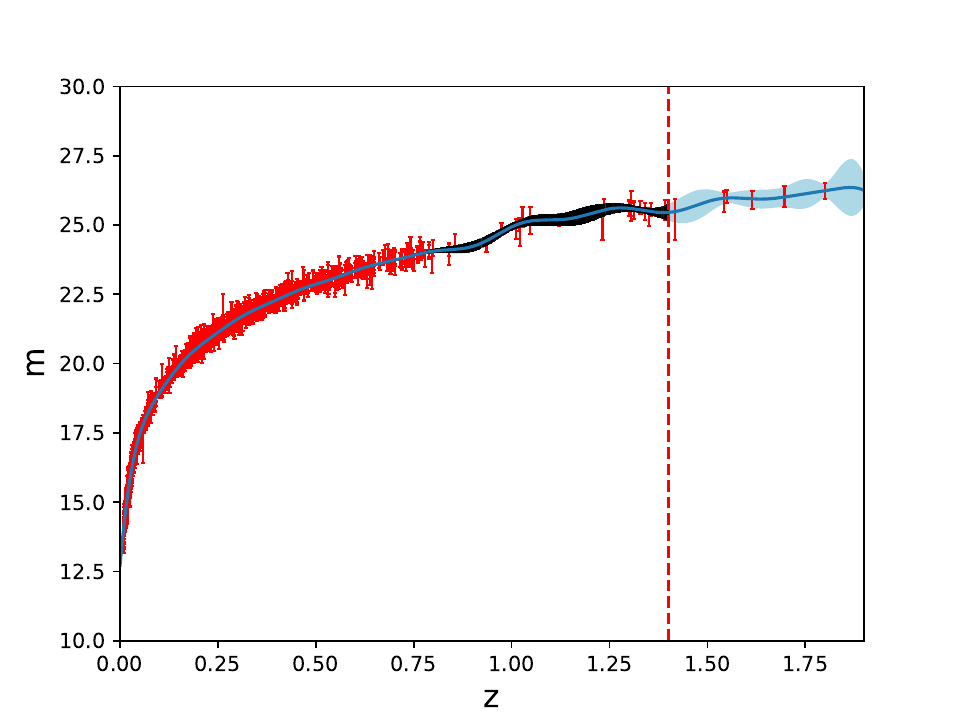}
	\caption{The blue curves depict the reconstructed function with $1\sigma$ uncertainty from the SN Ia data (red dots) through Gaussian process. % The black points denote the apparent magnitude of SN~Ia reconstructed at the redshift of the quasar.
	The red dashed line represents $z=1.4$.}
	\label{fig:restruction}
\end{figure}

To apply quasars in cosmology, we need to obtain their luminosity distances from the observables of quasars by using  the relations shown in Eqs.~(\ref{eq:std_F}, \ref{eq:evo_F}). Apparently, the values of the coefficients ($\alpha$, $\beta$, $\gamma$)  need to be determined first. Here, we adopt  the idea of distance ladder and  use the SN Ia data to calibrate the $L_X$-$L_{UV}$ relations. Since the data points of SN Ia concentrate in the low-redshift ($z\leq1.4$) region, the  SN Ia data from the latest Pantheon+ sample at $z\leq1.4$~\citep{Scolnic2022} are utilized to determine $\alpha$, $\beta$ and $\gamma$. 
To avoid the effect of cosmological models on the calibration, we use  the GP method~\citep{Seikel2012a} to construct the smooth curve of the apparent magnitude ($m$)-redshift relation of SN Ia.   GP is a generalization of Gaussian distribution, which refers to the distribution of random variables, and describes  the distribution over functions. With the GP method,   the smooth functions from a set of discrete data points can be constructed~\citep{Seikel2012a}. Here we use an infinitely differentiable squared exponential covariance function to reconstruct the smooth functions of the apparent magnitude of SN Ia. The covariance function is given by:
\begin{equation}
k(z, \tilde{z})=\sigma_{f}^{2} \exp \left[-\frac{(z-\tilde{z})^{2}}{2 l^{2}}\right],
\end{equation}
where the hyperparameters $\sigma_{f}$ and $l$ are optimized by maximizing the marginal likelihood \footnote{In this work, we use the public PYTHON package \textbf{GaPP} to reconstruct the apparent magnitude ($m$) -redshift relation of SN~Ia sample in low-redshift ($z<1.4$) region. This code is available at \href{https://github.com/astrobengaly/GaPP}{https://github.com/astrobengaly/GaPP}.}.
When the Pantheon+ sample~\citep{Scolnic2022}  is used,
we exclude those data whose redshifts are less than 0.01 since the nearby SN Ia sample may be impacted by the peculiar velocities~\citep{Brout2022}.
The reconstructed $m$-$z$ relation with an uncertainty of 1$\sigma$ by using the GP is shown in Fig.~\ref{fig:restruction}. This figure indicates that the reconstructed result is well consistent  with the data distribution.

\begin{figure}
	\centering
	\includegraphics[width=0.95\columnwidth]{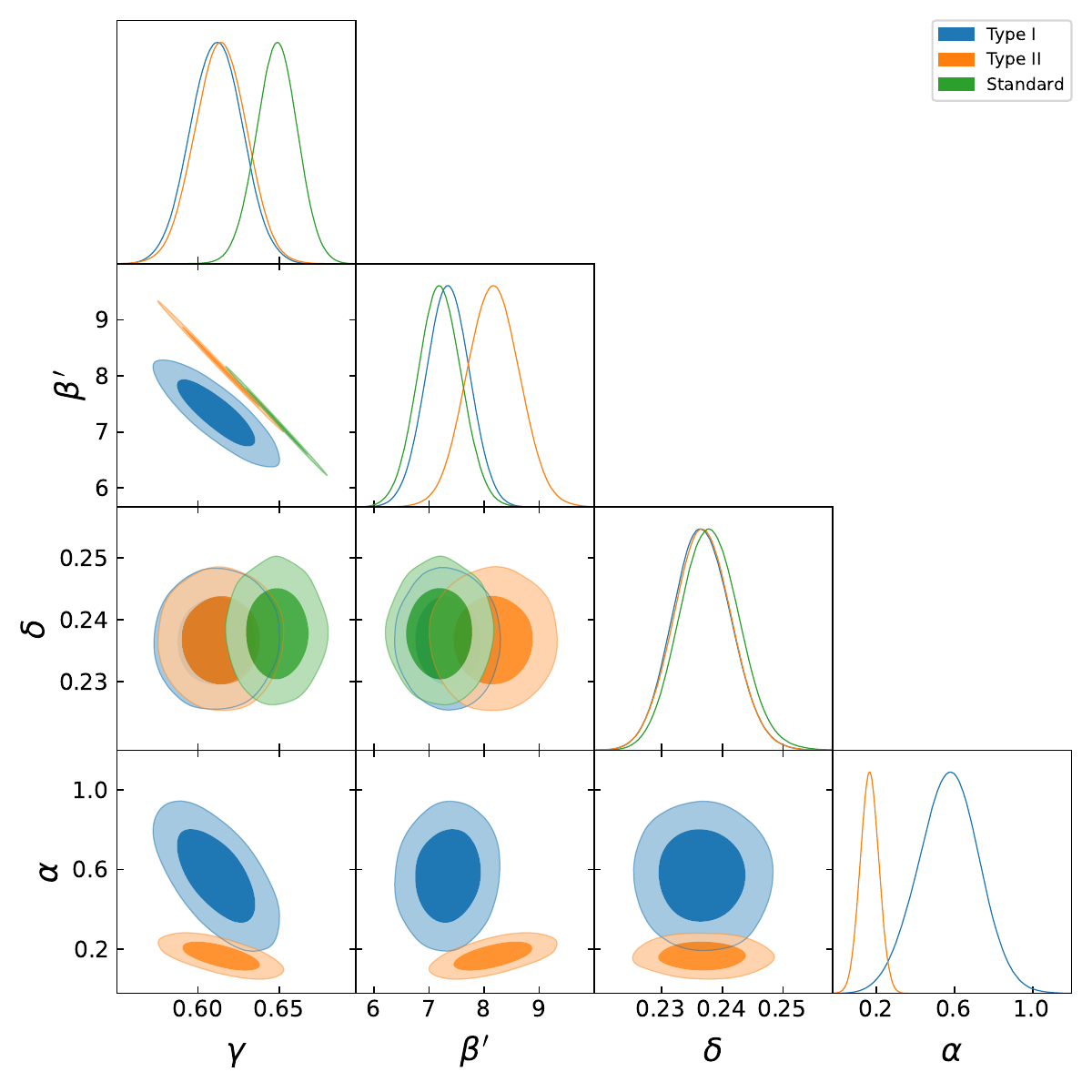}
	\caption{One-dimensional probability density distributions and  two-dimensional contours of $\alpha$, $\beta'$, $\gamma$, and $\delta$ from the low-redshift quasars. % with different $L_X$-$L_{UV}$ relations.
	The blue, orange, and green contours represent the Type I,  Type II, and  standard relations, respectively.}
	\label{fig:calibrate}
\end{figure}

The distance module $\mu$ of SN Ia  is related to the  luminosity distance ($d_{L}$) and the absolute magnitude~($M$) through
\begin{eqnarray}\label{eq:mu}
	\mu=m-M=5 \log\left(\frac{d_{L}}{\mathrm{Mpc}}\right)+25.
\end{eqnarray}
Thus, the luminosity distance of quasars that fall in the redshift under $z=1.4$ may be obtained from the reconstructed $m(z)$ after  the value of $M$ is determined.
However, the value of $M$ cannot be directly obtained using only the SN~Ia sample and thus $M$ is treated as a free parameter. Substituting Eq.~(\ref{eq:mu}) into the   $L_{X}$-$L_{UV}$ relations, one can obtain
\begin{align}\label{eq:evo_F_mu}
	\log \left(F_{X}\right)&=2(\gamma-1) \frac{m-M-25}{5}+\beta
	+(\gamma-1) \log (4 \pi)\nonumber \\
	&+\gamma \log \left(F_{U V}\right)+ \alpha \ln(\bar{\alpha}+z),
\end{align}
which shows  that there is a strong degeneracy between the absolute magnitude $M$ and the coefficient $\beta$.
Thus, we cannot constrain the values of $M$ and $\beta$ simultaneously by using the observational data.
To address this issue without assuming any prior values of $M$, we introduce a new coefficient: $\beta'=-2(\gamma-1)\left(\frac{M}{5}+5\right)+\beta+(\gamma-1)\log(4\pi)$. Then, the relation  Eq.~(\ref{eq:evo_F_mu}) is reduced to
\begin{eqnarray}\label{eq:evo_F_reduce}
	\log \left(F_{X}\right)  =2(\gamma-1) \frac{m}{5}+\beta'+\gamma \log \left(F_{U V}\right)+\alpha \ln(\bar{\alpha}+z).
\end{eqnarray}
 Using the apparent magnitude of SN~Ia reconstructed at a redshift of quasars and  the observed $\log \left(F_{U V}\right)_\mathrm{obs}$ at this redshift, we can obtain  $\log (F_{X})_\mathrm{re}$ from Eq.~(\ref{eq:evo_F_reduce}).
Comparing this $\log (F_{X})_\mathrm{re}$ with the observed X-ray flux of quasars can give the values of $\alpha$, $\beta'$ and $\gamma$  by maximizing the D'Agostinis likelihood function~\citep{D'Agostini2005}:
\begin{small}
\begin{eqnarray}\label{eq:likelihood_DA}
	&&\mathcal{L}(\delta,\bm{\theta} ) =  \prod_{i=1}^{N} \frac{1}{\sqrt{2\pi(\sigma^{2}_{i}+\delta^{2})}}  \times \\  \nonumber &&\exp \left\{-\frac{\left[\log \left(F_{X}\right)_{\mathrm{obs},i}-\log \left(F_{X}\right)_\mathrm{re}\left(\log \left(F_{UV}\right)_{\mathrm{obs},i}; \bm{\theta} \right)\right]^{2}}{2 (\sigma^{2}_{i}+\delta^{2})}\right\},
\end{eqnarray}
\end{small}
where $\delta$ is the intrinsic dispersion, $\bm{\theta}\equiv\left\{\alpha,~\beta',~\gamma \right\}$, and $\sigma_{i}^{2}=\sigma_{\log \left(F_{X}\right)_{\mathrm{obs},i}}^{2}+\gamma^{2} \sigma_{\log \left(F_{U V}\right)_{\mathrm{obs},i}}^{2}+\left(\frac{2 \gamma-2}{5}\right)^{2} \sigma_{m}^{2}$ with $\sigma_{\log \left(F_{X}\right)_{\mathrm{obs},i}}$, $\sigma_{\log \left(F_{U V}\right)_{\mathrm{obs},i}}$, and $\sigma_{m}$ being the uncertainties of $\log \left(F_{X}\right)_{\mathrm{obs},i}$, $\log\left(F_{UV}\right)_{\mathrm{obs},i}$, and the reconstructed $m$, respectively.
In Eq.~(\ref{eq:likelihood_DA}), number $N=1326$ is the number of quasars in the low-redshift ($z\leq 1.4$) region.
To find the maximum  likelihood, we utilize the Python package $emcee$~\citep{ForemanMackey2013}, which  bases on the Metropolis-Hastings algorithm and implements the Markov Chain Monte Carlo (MCMC) numerical fitting method.
The calibrated results for the coefficients in various $L_{X}$-$L_{UV}$ relations are presented in Tab.~\ref{tab1} and Fig.~\ref{fig:calibrate}.
Obviously, only the value of $\delta$ is almost independent of the  relations.  The values of $\gamma$ are almost the same in  the Type I and II relations, but they deviate from that in the standard $L_X$-$L_{UV}$ relation. The Type I and standard relations give almost the same value of $\beta'$, which is smaller that the one in the Type II relation. 
The Type I and II relations give different values of  $\alpha$, but both of them deviate from zero at more than $3\sigma$, which indicates that the observations seem to support  the redshift-evolutionary relations.

 To investigate the effect of different redshift borders,  we also  consider quasars in two different redshift regions ($z\leq1$ and $z\leq 1.8$). The results are shown in Tab.~\ref{tab2}.  Comparing Tab.~\ref{tab1} with Tab.~\ref{tab2}, we find that the relation coefficients and the intrinsic dispersions across these redshift regions vary slightly but remain consistent within a 1$\sigma$ CL. Thus, the choice of redshift borders exerts a negligible effect on  the calibration of the luminosity relations.  In the following analyses, we will use the results from quasars with $z\leq1.4$.

To further check whether there exist selection biases and redshift evolution in  quasar data~\citep{Singal2022,Singal2013,Singal2019}, we follow the process given in \citep{Dainotti2015,Dainotti2022} and use the EP method~\citep{Efron1992} to calculate the value of $k$ parameter which is introduced to correct the luminosity via $L_{\mathrm{corrected}}=L_{\mathrm{observed}}/(\bar{\alpha}+z)^k$ in Type I and II relations, and the corresponding value of $\alpha$. If the obtained value of $\alpha$ is consistent with that from the low-redshift calibration,  then the results presented in Table~\ref{tab1} are not influenced by selection biases or redshift evolution in quasars. When the correction to the luminosity is considered, we need to 
replace  $L_X$ and $L_{UV}$ with $L_X/(\bar{\alpha}+z)^{k_X}$ and $L_{UV}/(\bar{\alpha}+z)^{k_{UV}}$ in Eq.~(\ref{eq:std_L}).  Then, we obtain the following simple relation between $\alpha$ and $k_X$ and $k_{UV}$:
\begin{eqnarray}\label{eq:atok}
	\alpha=\frac{k_X-\gamma k_{UV}}{\ln10}.
\end{eqnarray}
Clearly, the value of $\alpha$ can be inferred from the values of $k_X$ and $k_{UV}$ when $\gamma$ is known.
The values of $k_X$ and $k_{UV}$ can be  obtained by using the EP method, which is given in a publicly available Mathematica code: \textit{Selection biases and redshift evolution in relation to cosmology}\footnote{\href{https://notebookarchive.org/2023-05-8b2lbrh}{https://notebookarchive.org/2023-05-8b2lbrh}}.
Here we set a fiducial model: the  flat $\Lambda$CDM model with $H_0=73.04~\mathrm{km~s^{-1}Mpc^{-1}}$ and  $\Omega_{\mathrm{m0}}=0.33$ for calculating the luminosity of quasars, and choose the flux limit $F_\mathrm{lim}=6\times10^{-33}~\mathrm{erg~s^{-1}cm^{-2}Hz^{-1}}$ for the X-rays and $F_\mathrm{lim}=4.5\times10^{-29}~\mathrm{erg~s^{-1}cm^{-2}Hz^{-1}}$ for the UV.
We finally obtain  $k_X=7.10\pm0.17$ and $k_{UV}=9.76\pm0.22$ for the Type I relation, and $k_X=2.64\pm0.06$ and $k_{UV}=3.62\pm0.08$ for the Type II relation.
Using the fiducial model  to estimate the luminosity distance, the value of parameter $\gamma$ can then be determined from  Eq.~(\ref{eq:likelihood_DA}) with the whole quasar sample, and the results are  $\gamma=0.579\pm 0.011$ and $\gamma=0.577\pm 0.011$ for the Type I and II relations, respectively.
Finally, we use Eq.~(\ref{eq:atok}) to calculate  the mean value of $\alpha$ with 1$\sigma$ uncertainty,  and obtain $\alpha=0.629\pm0.103$ and $\alpha=0.239\pm0.037$\footnote{This result is different from $0.342\pm 0.041$ obtained in \citep{Dainotti2022} since the luminosity is correced as $L'=L/g(z)$ with $g(z)=z^kz^k_{crit}/( z^k+z^k_{crit})$ rather than $g(z)=(1+z)^k$ in \citep{Dainotti2022}.} for the Type I and II relations, respectively.
These results deviate from zero at  more than $5\sigma$ and are  compatible with those  obtained from the low-redshift calibration and shown in Tab.~\ref{tab1}  within $2\sigma$ CL. This compatibility suggests that the nonzero value of  $\alpha$
 from the low-redshift calibration is unlikely to result from selection biases or redshift evolution in quasars.  It is worth noting that the bias due to extinction on quasar luminosity distances has been studied recently in \citep{Zajacek2024}.

\begin{table*}
\caption{Marginalized one-dimensional constraints on parameters with 1$\sigma$ CL from the low-redshift ($z \leq 1.4$) quasars\label{tab1}}
%\tablewidth{0pt}
\begin{tabular}{c|ccccc}
%\tablehead
\hline
	Relations & $\delta$ & $\alpha$ & $\beta'$ & $\gamma$ & $-2\ln\mathcal{L}$\\%&$\Delta$AIC & $\Delta$BIC
%\startdata
\hline
standard & $0.2379\pm0.0049$ & - & $7.19\pm0.40$ & $0.649\pm0.013$ & 0.939 \\%&11.209&6.109 \\
\hline
Type I & $0.2360^{+0.0046}_{-0.0051}$ & $0.57\pm 0.15$ & $7.33\pm 0.41$ & $0.612\pm 0.016$ & -12.270 \\%&-&- \\
\hline
Type II & $0.2370\pm 0.0049$ & $0.165\pm 0.046$ & $8.19\pm 0.48$ & $0.613\pm 0.016$ & -11.562 \\%&0.708&0.708 \\
%\enddata
\hline
\end{tabular}
\end{table*}

\begin{table*}
 \caption{Marginalized one-dimensional constraints on parameters with 1$\sigma$ CL from the quasars in two different redshift regions\label{tab2}}
 \renewcommand{\arraystretch}{1.5}
 %\tablewidth{0pt}
 \begin{tabular}{cccccc}
  %\tablehead
  \hline
  Redshift regions & Relations & $\delta$ & $\alpha$ & $\beta'$ & $\gamma$ \\
  %\startdata
  \hline
  & Standard & $0.2352\pm 0.0059$ & - & $7.45\pm 0.54$ & $0.640\pm 0.018$  \\
  \cline{2-6}
  $0<z\leq 1.0$ & Type I & $0.2350\pm 0.0061$ & $0.42^{+0.18}_{-0.24}$ & $7.18\pm 0.56$ & $0.625\pm 0.020$ \\
  \cline{2-6}
  & Type II & $0.2350\pm 0.0060$ & $0.111^{+0.048}_{-0.069}$ & $7.81\pm 0.59$ & $0.626\pm 0.019$ \\
  %\enddata
  \hline
  & Standard & $0.2358\pm 0.0042$ & - & $7.16\pm 0.32$ & $0.650\pm 0.010$  \\
  \cline{2-6}
  $0<z\leq 1.8$ & Type I & $0.2339\pm 0.0042$ & $0.51\pm 0.11$ & $7.59\pm 0.33$ & $0.607\pm 0.013$ \\
  \cline{2-6}
  & Type II & $0.2341\pm 0.0042$ & $0.165\pm 0.036$ & $8.39\pm 0.42$ & $0.607\pm 0.014$ \\
  %\enddata
  \hline
 \end{tabular}
\end{table*}

\section{Constraints on cosmological models}\label{sec3}
Extrapolating the values of $\alpha$, $\beta'$, $\gamma$ and $\delta$ from the low-redshift data to  the high-redshift regions, we can obtain the luminosity distance of all quasar data. These data  can now be employed to constrain the cosmological models by finding the maximum likelihood of the following function:
\begin{small}
\begin{eqnarray}\label{eq:likelihood}
 \tilde{\mathcal{L}} (\bm{p}) &=& \prod_{i=1}^{N} \frac{1}{\sqrt{2 \pi\left(\tilde{\sigma}_{i}^{2}+\delta^{2}\right)}} \times \\ \nonumber 
&&\exp \left\{-\frac{\left[\log \left(F_{X}\right)_{\mathrm{obs},i}-\log \left(F_{X}\right)_\mathrm{th}\left( %\log \left(F_{U V}\right)_{\mathrm{obs},i}; 
d_L(z_i, \bm{p})\right)\right]^{2}}{2\left(\tilde{\sigma}_{i}^{2}+\delta^{2}\right)}\right\}.
\end{eqnarray}
\end{small}
Here $N$ denotes the number of data, and $\tilde{\sigma}_{i}$ is the derived uncertainty of $\log \left(F_{X}\right)$ by using the error propagation formula
\begin{small}
\begin{eqnarray}
	\tilde{\sigma}_{i}^{2}&=&\sigma_{{\log \left(F_{X}\right)}_{\mathrm{obs},i}}^{2}+\left(\frac{\partial \log \left(F_{X}\right)_\mathrm{t h}}{\partial \gamma}\right)_{i}^{2} \sigma_{\gamma}^{2} \nonumber\\ &+&\left(\frac{\partial \log \left(F_{X}\right)_\mathrm{t h}}{\partial \beta'}\right)_{i}^{2} \sigma_{\beta'}^{2}
	+\left(\frac{\partial \log \left(F_{X}\right)_\mathrm{t h}}{\partial \alpha}\right)_{i}^{2} \sigma_{\alpha}^{2} \nonumber\\ &+&\left(\frac{\partial \log \left(F_{X}\right)_\mathrm{t h}}{\partial \log \left(F_{U V}\right)_\mathrm{obs}}\right)_{i}^{2} \sigma_{\log \left(F_{U V}\right)_\mathrm{obs}}^{2}\nonumber\\
	&+&2 \sum_{j=1}^3 \sum_{k=j+1}^3\left(\frac{\partial \log \left(F_{X}\right)_\mathrm{t h}}{\partial {\theta}_{j}} \frac{\partial \log \left(F_{X}\right)_\mathrm{t h}}{\partial {\theta}_{k}}\right)_{i} C_{j k},
\end{eqnarray}
\end{small}
where $C_{j k}$ is the covariance matrix, and $\sigma_\alpha$,
$\sigma_{\beta'}$, and $\sigma_\gamma$ are the uncertainties of coefficients $\alpha$, $\beta'$, and $\gamma$ estimated from the GP, respectively.
In Eq.~(\ref{eq:likelihood}), $\bm{p}$ represents the cosmological model parameters and $m$ is the apparent magnitude derived  from the cosmological model, which relates to  the dimensionless luminosity distance $D_L$ through
\begin{eqnarray}\label{eq:mth}
	m(z, \bm{p})=5 \log D_{L}(z;\bm{p}) + \mathcal{M},
\end{eqnarray}
with
\begin{eqnarray}
	\mathcal{M}=25 + M - 5\log \frac{c}{H_0}.
\end{eqnarray}
The dimensionless luminosity distance  is defined as
\begin{eqnarray}\label{eq:dL}
	D_L(z;\bm{p})=\frac{H_0}{c} d_L(z;\bm{p})=  (1+z) \int_{0}^{z}\frac{d\tilde{z}}{E(\tilde{z};\bm{p})}
\end{eqnarray}
in the spatially flat universe. Here, $E({z};\bm{p})$ is the dimensionless Hubble parameter.
For the $w$CDM model, $E(z;\bm{p})$ has the form
\begin{eqnarray}
	E(z;\bm{p})^2=\Omega_{\mathrm{m0}}(1+z)^3+(1-\Omega_{\mathrm{m0}})(1+z)^{3(1+w)},
\end{eqnarray}
where $\Omega_{\mathrm{m0}}$ and $w$ are the current matter density and the equation of state (EOS) of dark energy, respectively.
This model will reduce to the  $\Lambda$CDM when $w=-1$. Therefore, we have $\bm{p}\equiv\left\{\Omega_{\mathrm{m0}}\right\}$ for the $\Lambda$CDM model and $\bm{p}\equiv\left\{w,\Omega_{\mathrm{m0}}\right\}$ for the $w$CDM model. 
Here the prior of $w$ is set as a uniform distribution in $1/(\Omega_{\mathrm{m0}}-1)\leq w\leq -1/3$, which comes from the accelerated cosmic expansion \citep{Riess1998, Perlmutter1999} and the null energy  condition~\citep{Visser2000,Lenart2023}.
Since  $M$ and $H_0$ can not be constrained simultaneously when only quasars are used, 
$\mathcal{M}$ is treated as a nuisance parameter, and is  marginalized in our analysis.

\subsection{Constraints on cosmological models from Quasars}

\begin{figure}
	\centering
	\includegraphics[width=.48\columnwidth]{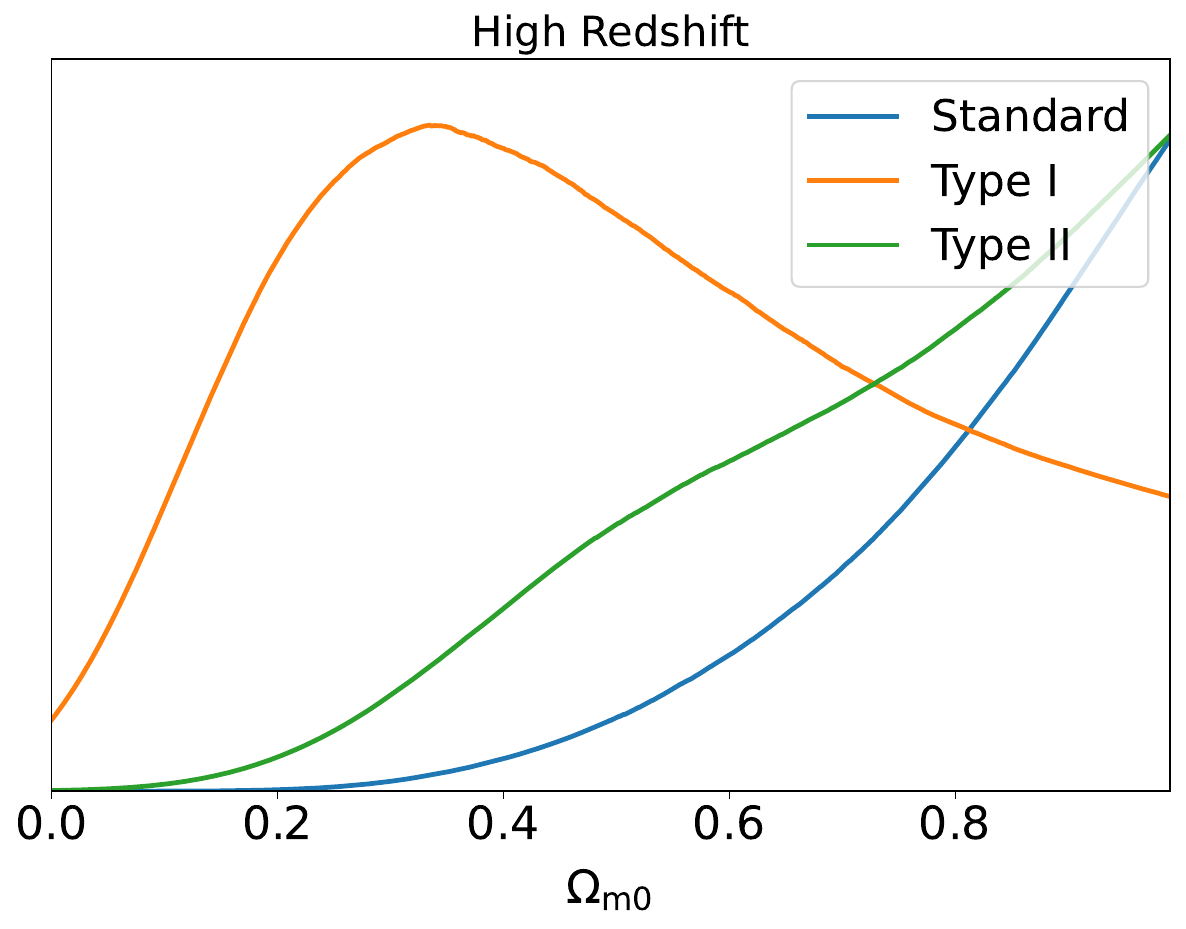}
	\includegraphics[width=.48\columnwidth]{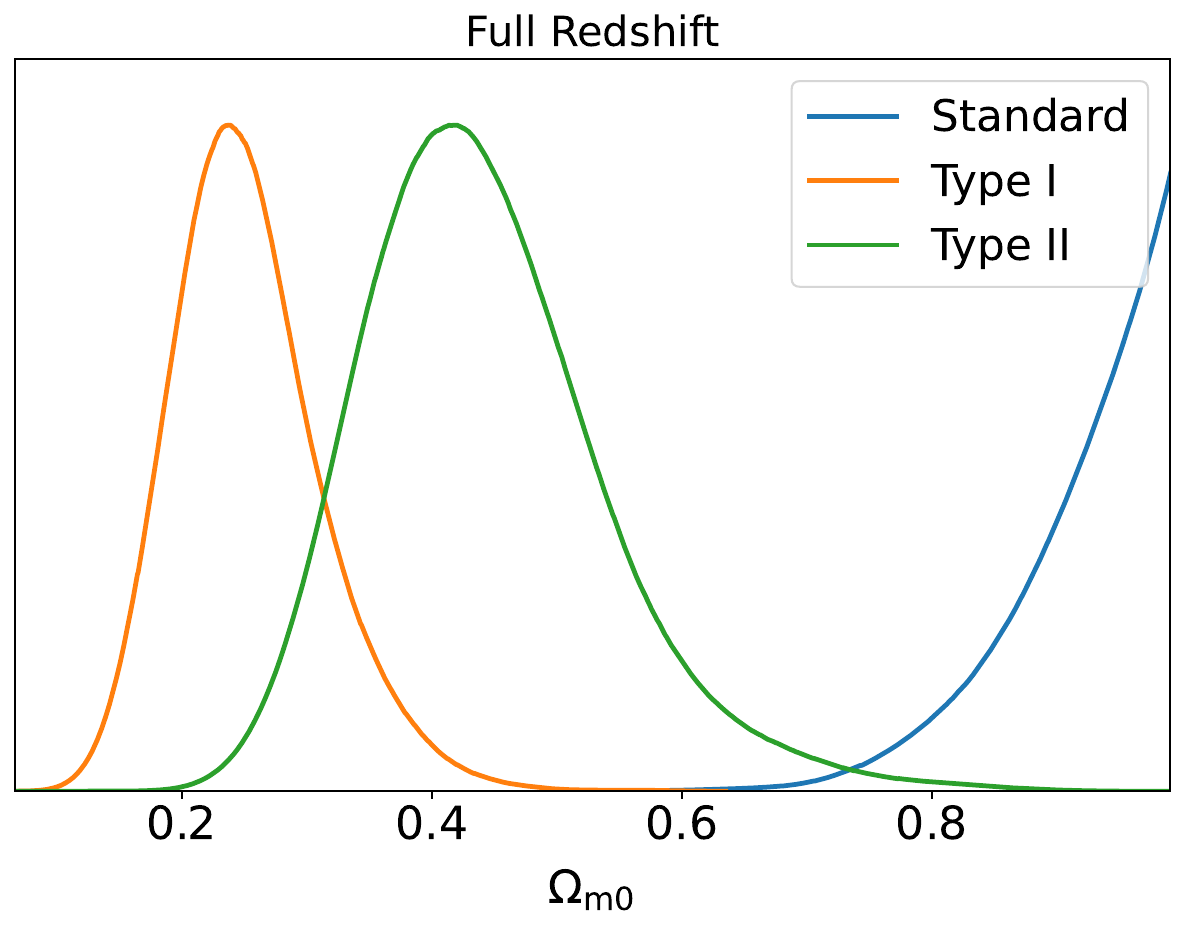}
	\caption{
		The marginalized probability density distributions of  $\Omega_{\mathrm{m0}}$ in the $\Lambda$CDM model, obtained from high-redshift ($z>1.4$) and full-redshift quasars, are analyzed for three different $L_{X}$-$L_{UV}$ relations.
	}
	\label{fig:LCDM}
\end{figure}

\begin{figure}
	\centering
\includegraphics[width=.48\columnwidth]{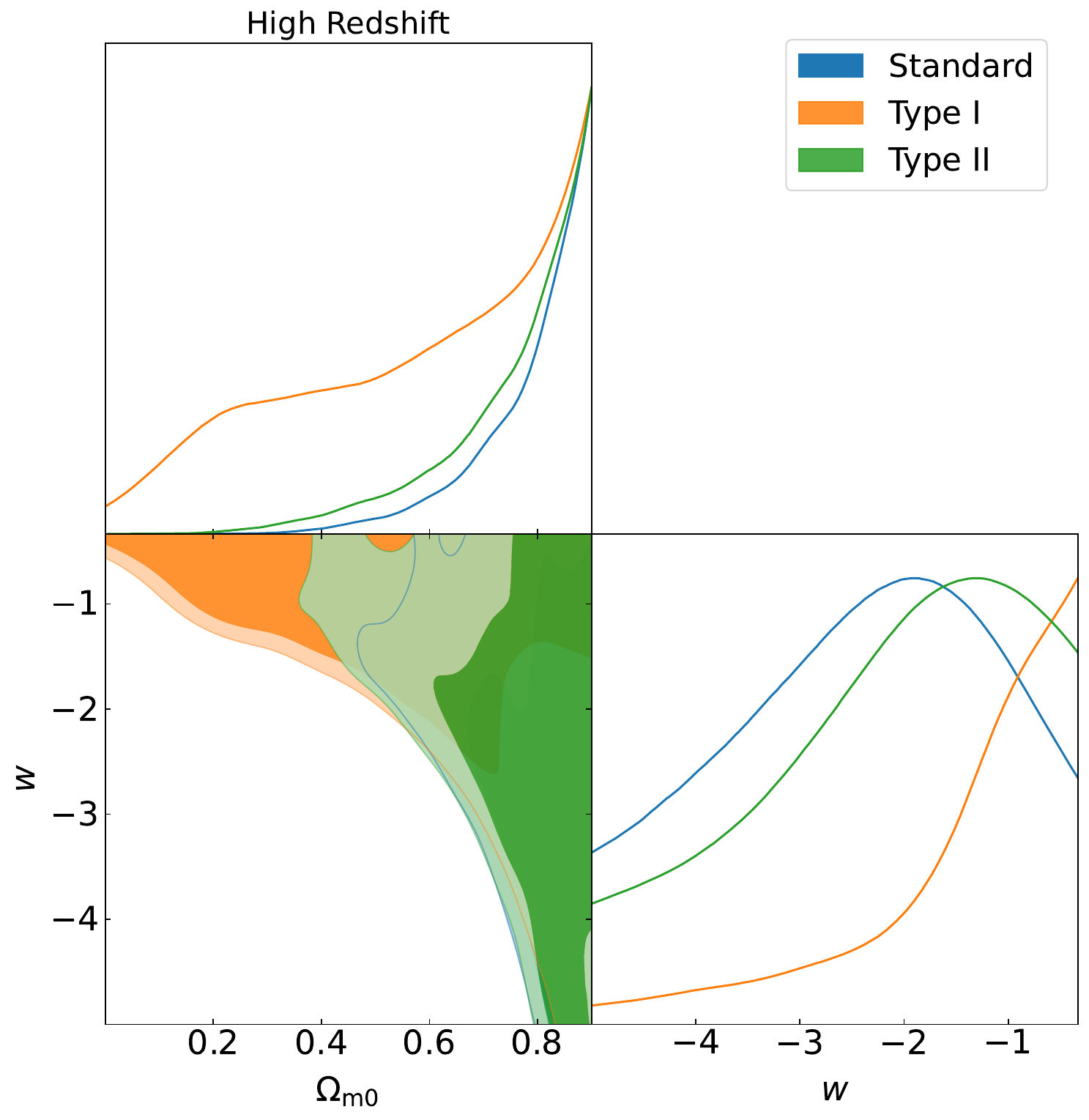}
\includegraphics[width=.48\columnwidth]{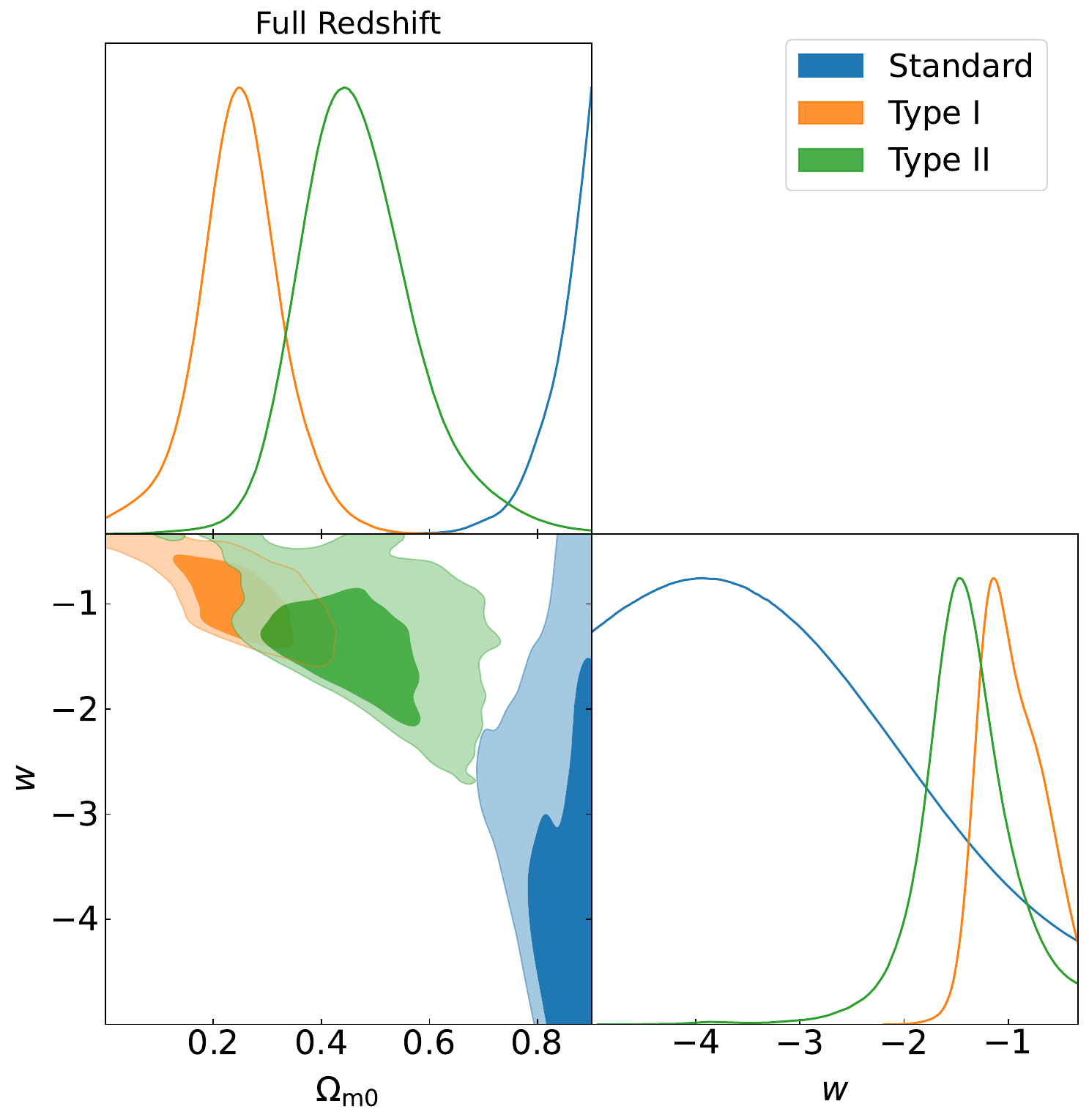}
	\caption{
		Constraints on  $\Omega_{\mathrm{m0}}$ and $w$ in the $w$CDM model, obtained from high-redshift ($z>1.4$) and full-redshift quasars, are analyzed for three different $L_{X}$-$L_{UV}$ relations.
	}
	\label{fig:wCDM}
\end{figure}

In Sec. II, we have used the SN Ia data in the redshift region of $z\leq 1.4$ to calibrate the quasars in this redshift region and extrapolate the results to the high-redshift ($z>1.4$) quasars. Thus, the high-redshift quasars, which contain 1095 observations,  are utilized firstly to constrain the $\Lambda$CDM and $w$CDM models. Furthermore, we also consider the full-redshift quasar data (2421 data points) to limit these two models. The results are shown in Figs. \ref{fig:LCDM} and \ref{fig:wCDM}, and summarized in  Tab. \ref{tab3}.

It is easy to see that  for the $\Lambda$CDM model when the standard $L_X$-$L_{UV}$ relation is used, quasars only  give a lower bound on $\Omega_\mathrm{m0}$. The full-redshift quasars with Type II relation give an effective constraint on $\Omega_{\mathrm{m0}}$, while the high-redshift ones do not. When Type I  relation is used, both the full- and high-redshift quasars can constrain  $\Omega_{\mathrm{m0}}$ effectively. For the $w$CDM model,   the effective constraints on $\Omega_{\mathrm{m0}}$ and $w$ can be achieved only from the full-redshift quasars with  Type I or II relation.  In both the $\Lambda$CDM and $w$CDM models,  quasars with  Type I relation always tend to give smaller $\Omega_{\mathrm{m0}}$ and $-2\ln\mathcal{L}$  than those with the  standard and Type II relations.

\begin{figure}
	\centering	\includegraphics[width=.48\columnwidth]{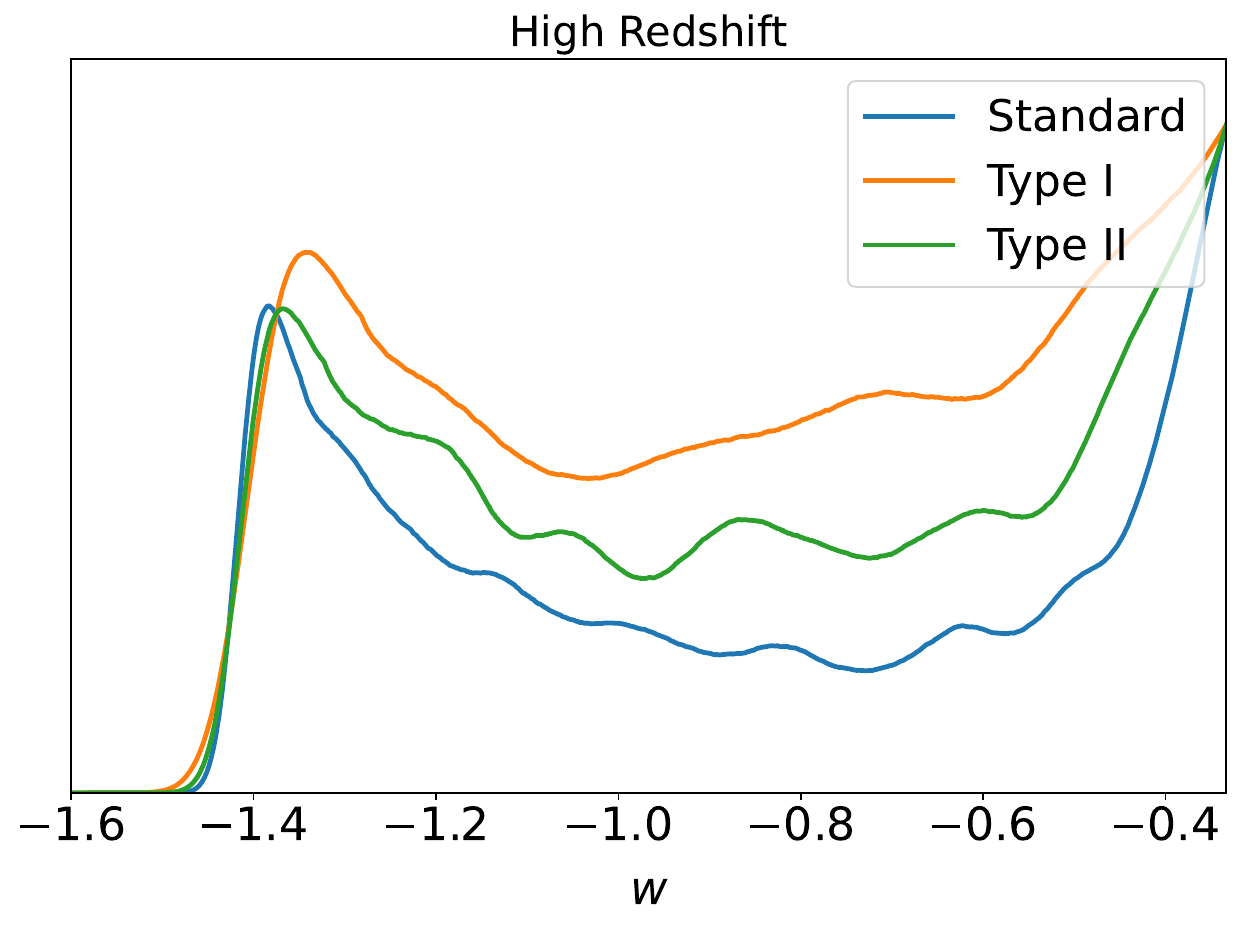}	\includegraphics[width=.48\columnwidth]{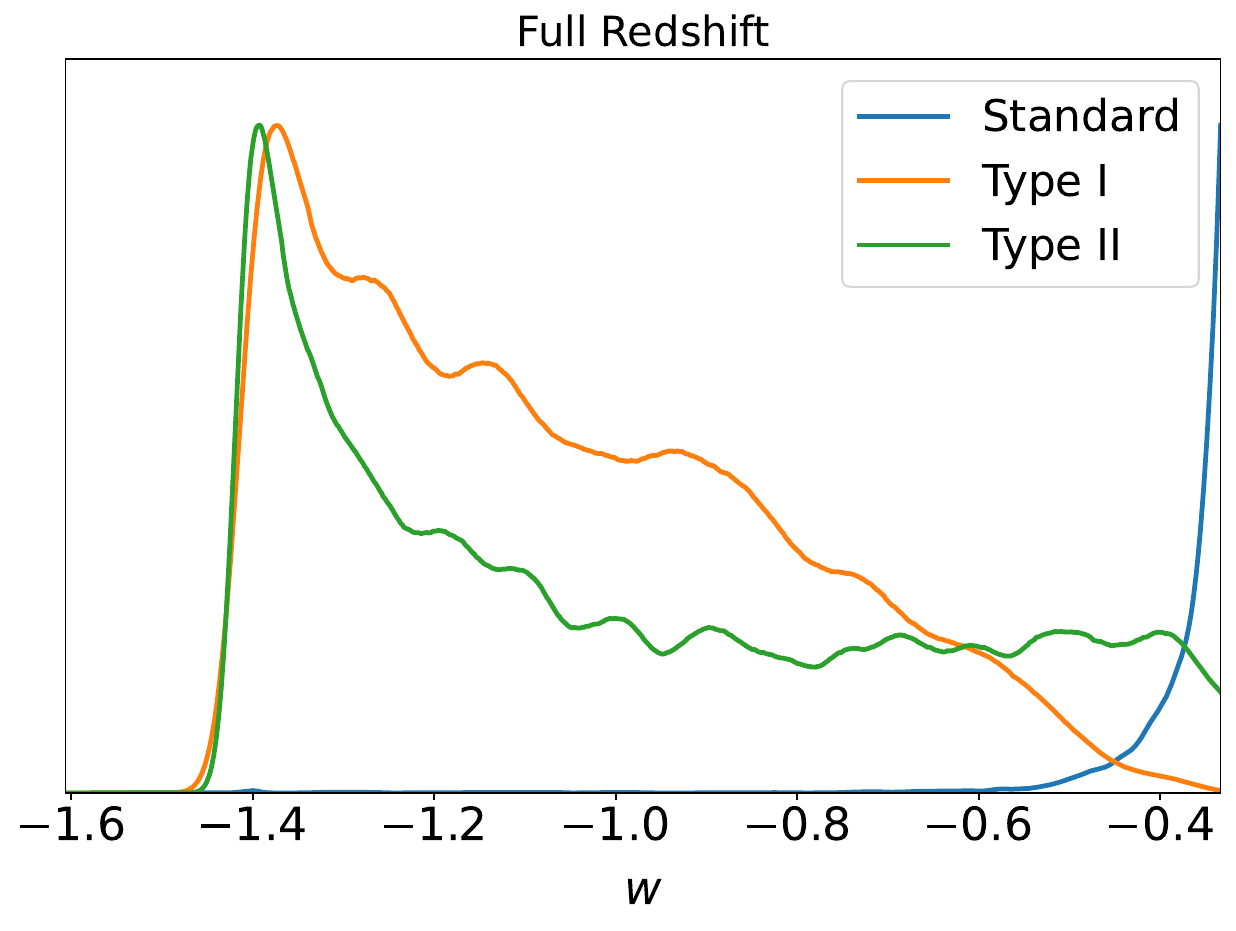}
	\caption{
		Constraints on $w$ with  $\Omega_{\mathrm{m0}}=0.30$ in the $w$CDM model  from high-redshift ($z>1.4$) and full-redshift quasars.
	}
	\label{fig:wCDM_omm}
\end{figure}

Since the cosmological  parameters $\Omega_{\mathrm{m0}}$ and $w$ cannot be well constrained simultaneously, we further consider a fitting for  $w$ with  $\Omega_{\mathrm{m0}}$ fixed to be $0.30$.
The results are shown in Fig.~\ref{fig:wCDM_omm}.
Apparently, for all three relations the high-redshift quasars only give the lower limit of $w$.  When the full-redshift quasars are used, we obtain that $w=-1.08^{+0.17}_{-0.34}$ for Type I relation and $w=-1.02^{+0.63}_{-0.41}$ for Type II relation.  If the standard relation is considered,  a large $w$ ($-0.381<w<-1/3$) is obtained.

\subsection{Constraints on cosmological models from   Quasars + OHD}

Since  quasars cannot provide the constraints on $M$ and $H_0$ due to the degeneracy between them, we further add the OHD from the cosmic age difference method~\citep{Jimenez2002}  into our analysis.
The updated 32 OHD measurements cover a redshift range of $0.07<z<1.965$ \citep{Moresco2020}, which contains 17 uncorrelated and 15 correlated measurements. The 17 uncorrelated data are given in~\citep{Zhang2014, Simon2005,Ratsimbazafy2017} and the 15 correlated measurements are sourced from \citep{Moresco2012,Moresco2015,Moresco2016} with the covariance matrix being  given by \cite{Moresco2020}.
To constrain the cosmological models, the minimization of the $\chi^2$ method is used:
\begin{equation}\label{eq:chi_Hz}
\chi_\mathrm{OHD}^{2}(\bm{p})=\sum_{i=1}^{17} \frac{\left[H_\mathrm{th}(z_i;\bm{p})-H_{\mathrm{obs}}(z_{i})\right]^{2}}{\sigma_{H, i}^{2}}+\Delta \hat{H}^{T} \mathbf{C}_{H}^{-1} \Delta \hat{H}.
\end{equation}
Here $\sigma_{H}$ represents the observed uncertainty of 17 uncorrelated measurements, $\Delta\hat{H}=H_\mathrm{th}(z;\bm{p})-H_{\mathrm{obs}}(z)$ represents the difference vector between the observed data and the theoretical values for the 15 correlated measurements, and $\mathbf{C}_{H}^{-1}$ is the inverse of the covariance matrix. $H_\mathrm{obs}$ represents the observed value of the Hubble parameter, while $H_\mathrm{th}(z;\bm{p})\equiv H_0E(z;\bm{p})$  is the corresponding  theoretical one calculated from the cosmological model.

\begin{figure}
	\centering
	\includegraphics[width=.48\columnwidth]{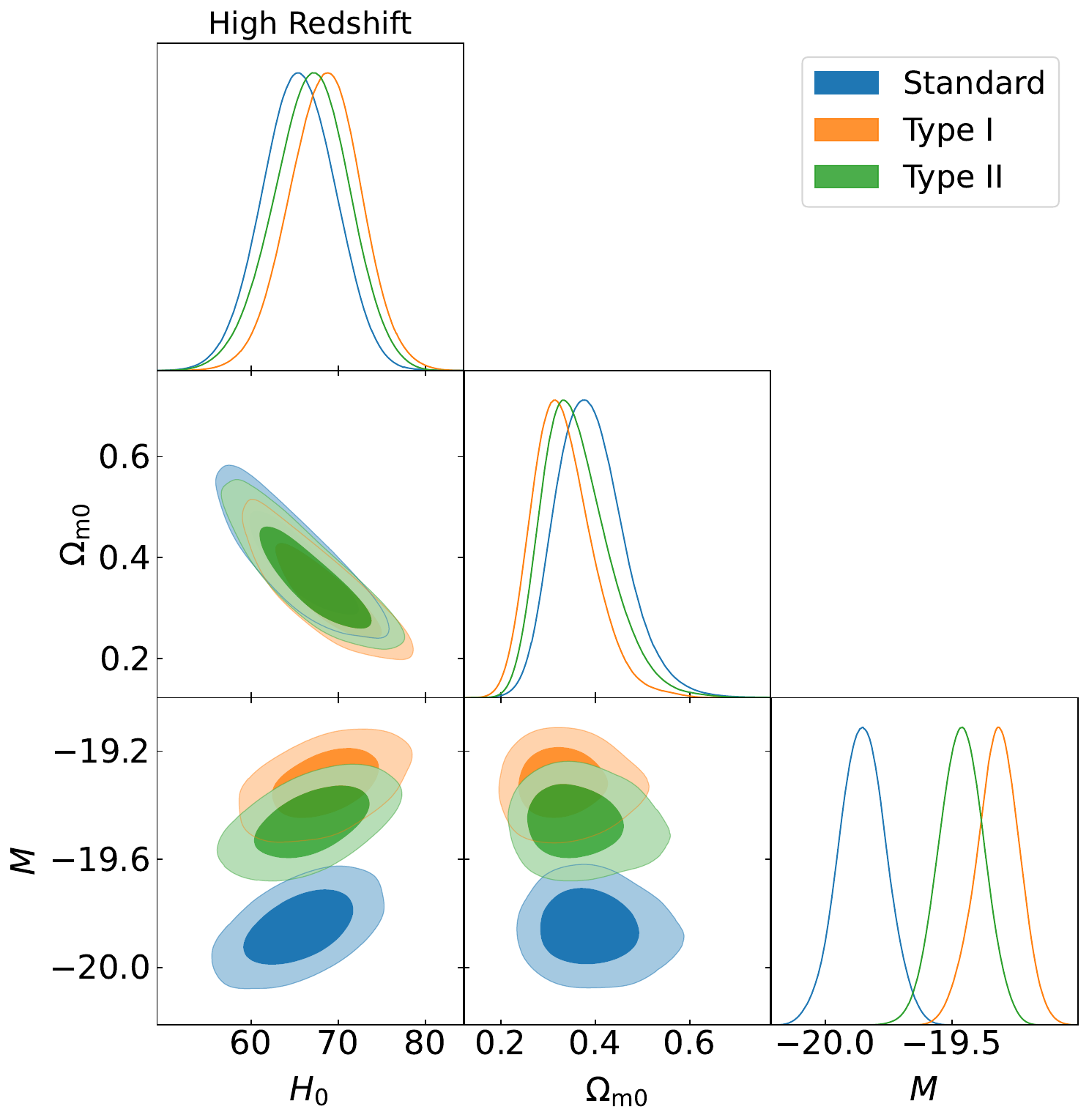}
	\includegraphics[width=.48\columnwidth]{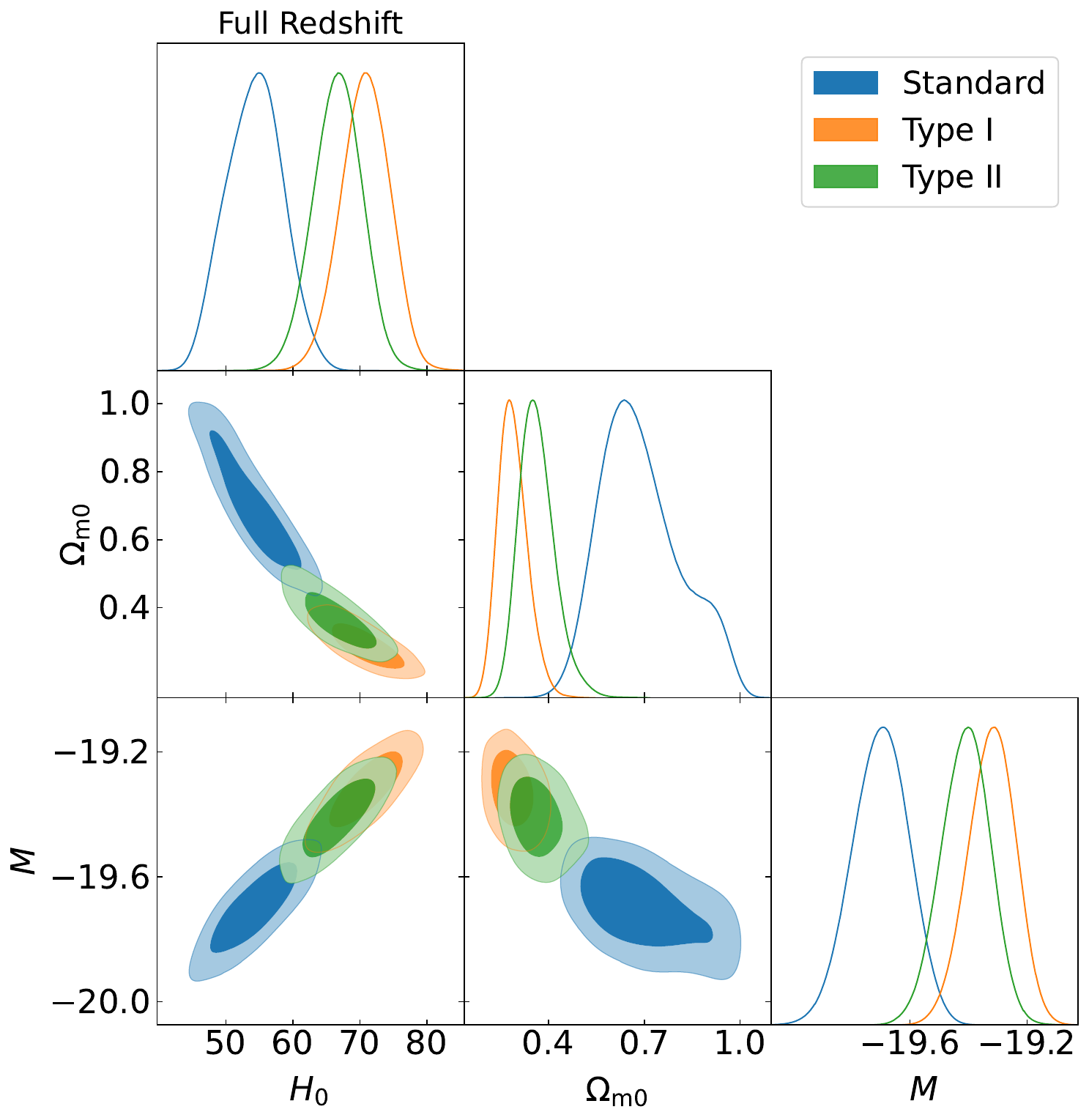}
	\caption{
		Constraints on $H_0$, $\Omega_{\mathrm{m0}}$ and $M$ in the $\Lambda$CDM model  for three different $L_{X}$-$L_{UV}$ relations, obtained from high-redshift ($z>1.4$) and full-redshift quasars plus OHD.
	}
	\label{fig:LCDM_OHD}
\end{figure}

\begin{figure}
	\includegraphics[width=.48\columnwidth]{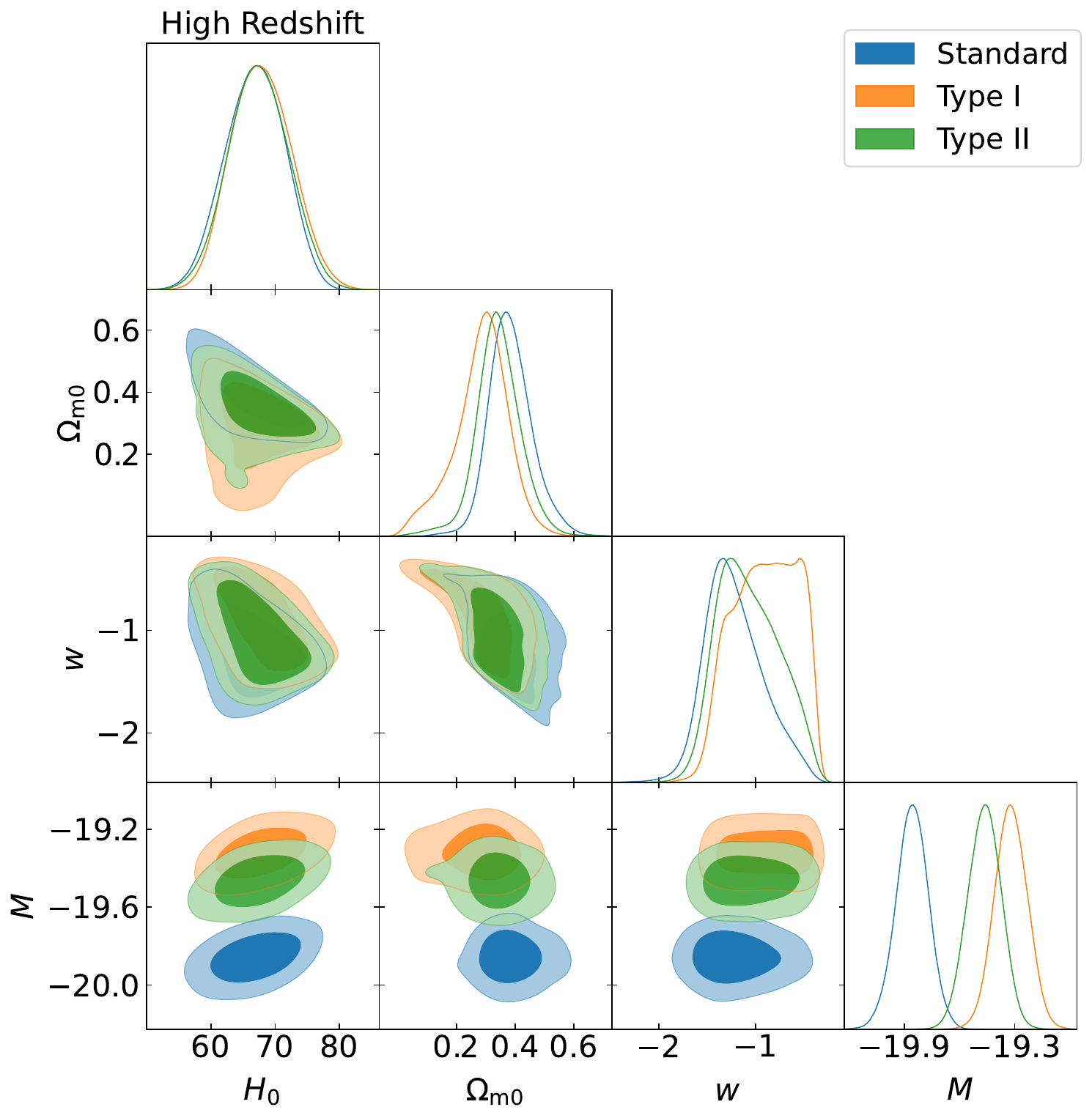}
	\includegraphics[width=.48\columnwidth]{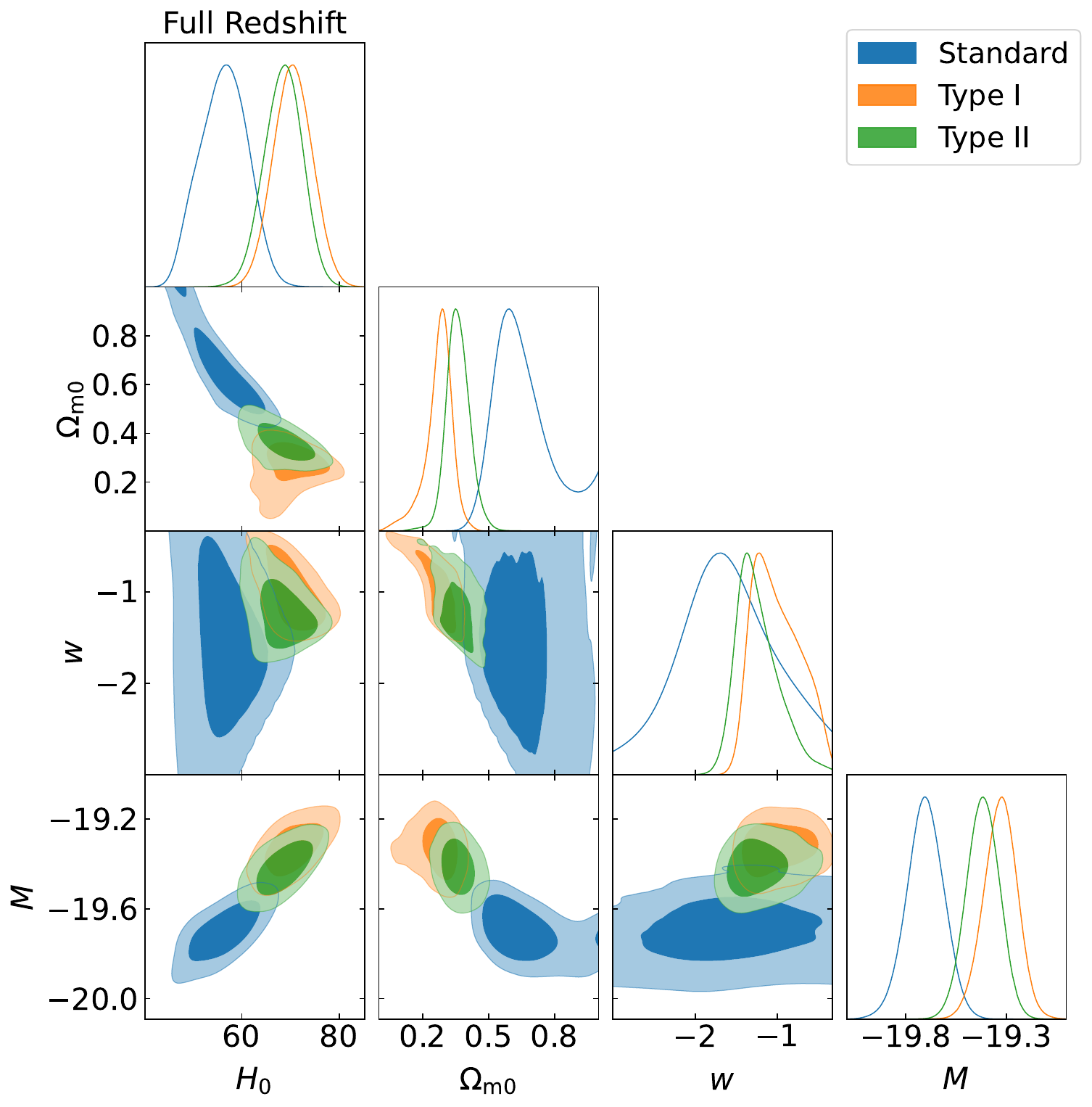}
	\caption{
		Constraints on $H_0$, $\Omega_{\mathrm{m0}}$, $w$ and $M$ in the $w$CDM model  for three different $L_{X}$-$L_{UV}$ relations, obtained from high-redshift ($z>1.4$) and full-redshift quasars plus OHD.
	}
	\label{fig:wCDM_OHD}
\end{figure}

The constraints on the cosmological models from quasars+OHD can be obtained from the total likelihood:
\begin{equation}
	\ln\mathcal{L}_\mathrm{total} (\bm{p})=  \ln\tilde{\mathcal{L}}(\bm{p})-\frac{1}{2}\chi_\mathrm{OHD}^{2}(\bm{p}).
\end{equation}
The results are shown in Figs.~\ref{fig:LCDM_OHD} and \ref{fig:wCDM_OHD}, and summarized in Tab.~\ref{tab4}.

For the $\Lambda$CDM model,  we find that in the case of the standard $L_X$-$L_{UV}$ relation the combination of the full-redshift quasars and the OHD measurements favors a smaller $H_0$, a larger $\Omega_\mathrm{m0}$, and a larger $M$ than the high-redshift quasars plus  OHD. The differences between the values of the cosmological model parameters and the absolute magnitude respectively from the high- and full-redshift quasars found in the standard relation become negligible when the Type II or Type I relation is used instead.   In the Type II relation, the value of $H_0$ is consistent with the result from the Planck CMB observations~\citep{Planck2020}, while the value of $M$ is smaller than the SH0ES measurement~\citep{Riess2022}. 
When the Type I relation is considered, we find that the $H_0$ value from full-redshift quasars locates between the Planck result and the SH0ES measurement~\citep{Riess2022,Planck2020}, and the $M$ value is compatible with the SH0ES measurement~\citep{Riess2022}.

For the $w$CDM model, we find that there is a large difference between the constraints on $H_{0}$ obtained from the high- and full-redshift quasars in the standard  $L_X$-$L_{UV}$ relation, and this difference reduces to be very small when the Type I or II relation is considered. 
For the standard relation,  the full-redshift quasars favor a smaller $H_0$ than the high-redshift data, but the results obtained in the Type I and II relations are opposite.  For  Type I and  II relations, the constraints on $\Omega_\mathrm{m0}$ obtained from the high- and full-redshift quasars are very close, while the values obtained from Type I relation are always smaller than those obtained from  Type II one.
The standard relation gives two significantly different $\Omega_{\mathrm m0}$  from the full- and high-redshift quasars, respectively, which are larger than those obtained in  Type I and II relations.
In the three relations,  quasars can always give  constraints on $w$ consistent with -1 at 2$\sigma$ CL.
The constraints on $M$ obtained from both the high- and full-redshift quasars in these three relations are nearly identical to that obtained in the $\Lambda$CDM model.  Furthermore, quasars with the Type I relation have the minimum  value of $-2\ln\mathcal{L}$.

\begin{table*}
\centering
\caption{ Constraints on the $\Lambda$CDM and $w$CDM models from Quasars. \label{tab3}}
\begin{tabular}{c|c|cccc}

\hline
		Relation & Model & Data Set &$\Omega_{\mathrm{m0}}$  & $w$ &$-2\ln\mathcal{L}$\\ \hline
	Standard & $\Lambda$CDM &$z>1.4$  & $> 0.765$ &-& -94.716 \\
	& & full~$z$  & $> 0.903$ &-& -95.963 \\
	\hline
	Type I & $\Lambda$CDM &$z>1.4$  &$0.49^{+0.21}_{-0.33}$ &-& -129.022 \\
	& & full~$z$ & $0.253^{+0.046}_{-0.067}$ &-& -153.029 \\
	\hline
	Type II & $\Lambda$CDM &$z>1.4$  & $> 0.622$ &-& -129.191 \\
	& & full~$z$ &  $0.440^{+0.073}_{-0.11}$ &-& -147.054  \\
	\hline \hline
	Standard & $w$CDM & $z>1.4$ & $> 0.766$ & $-3.23^{+2.8}_{-0.70}$& -94.156 \\
	& & full~$z$ & $> 0.834$ & $-4.4^{+2.3}_{-1.8}$& -94.305  \\
	\hline
	Type I & $w$CDM & $z>1.4$ &  $> 0.468$ & $> -2.08$ & -129.631  \\
	& & full~$z$ & $0.249\pm 0.082$ & $-0.98^{+0.25}_{-0.34}$ & -153.638 \\
	\hline
	Type II & $w$CDM & $z>1.4$  & $> 0.724$ & $> -3.30$& -129.057  \\
	& & full~$z$ & $0.469^{+0.084}_{-0.12}$ & $-1.43\pm 0.45$ & -149.635 \\
	\hline
		\end{tabular}\\
		{Note$-$The marginalized mean values, the standard deviations, and the $68\%$ CL.}
	%\tablecomments{The marginalized mean values, the standard deviations, and the $68\%$ CL.}
\end{table*}

\begin{table*}
\caption{ Constraints on the $\Lambda$CDM and $w$CDM models from quasars and OHD. \label{tab4}}
%	\tablewidth{0pt
\small{
\begin{tabular}{c|c|cccccc}
%	\tablecaption{ Constraints on the $\Lambda$CDM and $w$CDM models from quasars and OHD. \label{tab3}}
%	\tablewidth{0pt}
%	\small{
	%\tablehead{
	\hline
		Relation & Model & Data Set& $H_{0} $ &$\Omega_{\mathrm{m0}}$ & $w$& $M$ &$-2\ln\mathcal{L}$ \\
		\hline
	%}
	%\startdata
	Standard & $\Lambda$CDM &$z>1.4$  & $65.5\pm 4.2$ &$0.390^{+0.059}_{-0.081}$&-&$-19.854\pm 0.093$& -72.484  \\
    & & full $z$  & $54.3\pm 4.2$ &$0.693^{+0.088}_{-0.16}$&-&$-19.700\pm 0.096$& -56.918 \\
    \hline
    Type I & $\Lambda$CDM &$z>1.4$  & $68.5\pm 4.1$ &$0.332^{+0.050}_{-0.073}$&-&$-19.321^{+0.090}_{-0.079}$& -114.503 \\
    & & full $z$ & $70.8\pm 3.6$ &$0.289^{+0.038}_{-0.051}$&-&$-19.321^{+0.085}_{-0.076}$& -137.804 \\
    \hline
    Type II & $\Lambda$CDM &$z>1.4$  & $66.9\pm 4.3$ &$0.361^{+0.052}_{-0.083}$&-&$-19.464\pm 0.090$& -111.631 \\
    & & full $z$ & $66.7\pm 3.6$ &$0.364^{+0.045}_{-0.063}$&-&$-19.410\pm 0.083$& -131.713 \\ \hline \hline
	Standard & $w$CDM & $z>1.4$  & $66.9\pm 4.6$&$0.384^{+0.061}_{-0.078}$&$-1.22^{+0.24}_{-0.35}$&$-19.858\pm 0.089$& -73.892 \\
    & & full $z$ & $56.0^{+5.5}_{-4.8}$&$0.658^{+0.076}_{-0.16}$&$-1.69^{+0.67}_{-0.75}$&$-19.702\pm 0.092$& -58.194  \\
    \hline
    Type I & $w$CDM & $z>1.4$ &  $67.9^{+4.4}_{-5.0}$&$0.283^{+0.10}_{-0.070}$&$-0.89^{+0.45}_{-0.25}$&$-19.320\pm 0.089$& -114.503  \\
    & & full $z$ & $70.5\pm 4.0$&$0.272^{+0.067}_{-0.040}$&$-1.01^{+0.22}_{-0.37}$&$-19.330\pm 0.083$& -138.321  \\
    \hline
    Type II & $w$CDM & $z>1.4$  &$67.5\pm 4.6$&$0.342\pm 0.079$&$-1.07^{+0.27}_{-0.39}$&$-19.465\pm 0.091$& -112.232  \\
    & & full $z$ & $68.5\pm 3.9$&$0.360^{+0.048}_{-0.054}$&$-1.24^{+0.18}_{-0.31}$&$-19.416\pm 0.081$& -133.843 \\
	\hline
	%\enddata
	%}
\end{tabular}}\\
{Note$-$The marginalized mean values, the standard deviations, and the $68\%$ CL.}
\end{table*}

\section{discussion}
\label{sec4}

Except for the low-redshift calibration, the global fitting method is usually used to avoid the so-called circularity problem. Using the total 2421 quasar datapoints to constrain the spatially flat $\Lambda$CDM model and the coefficients in the $L_X$-$L_{UV}$ relations simultaneously, it has been found that $\Omega_{\mathrm{m0}}>0.924$, $\delta=0.228\pm{0.004}$, $\beta=7.021\pm0.249$ and $\gamma=0.639\pm0.008$ for the standard relation, and $\Omega_{\mathrm{m0}}=0.510^{+0.163}_{-0.254}$, $\delta=0.225\pm{0.004}$, $\beta=7.825\pm0.316$, $\gamma=0.579\pm0.011$ and $\alpha=0.580^{+0.084}_{-0.099}$ for Type I relation~\citep{Wang2022}. Apparently, the values of relation coefficients and intrinsic dispersion from the global fitting are well consistent with  those shown Tab.~\ref{tab1}, and the values of $\Omega_{\mathrm{m0}}$ are compatible with those (see Tab.~\ref{tab3}) from the high-redshift quasars. In the flat $\Lambda$CDM model, quasars in the redshift region $0.0041 \leq z \leq 1.686$ give that the best fitting values are  $\Omega_{\mathrm{m0}}=0.995$, $\delta=0.329$ and $\gamma=0.559$~\citep{Khadka2023}. These values of $\delta$ and $\gamma$ are compatible with those shown in Tab.~\ref{tab1}. 
Assuming a cosmological model with parameter values based on the computations with SNe Ia, \cite{Lenart2023} obtained that $\delta=0.231\pm0.004$,  $\beta=6.817\pm0.265$, $\gamma=0.648\pm0.009$ for the standard relation, and  $\delta=0.231\pm0.004$,  $\beta=8.278\pm0.362$, $\gamma=0.591\pm0.013$ for the Type II relation with $\alpha$ fixed from the EP method.  These results are also consistent with ours obtained from the low redshift calibration. Comparing Fig.~\ref{fig:calibrate} with Fig.~2a and 2b in \citep{Dainotti2023c} and Figs.~2 and 3 in \citep{Bargiacchi2023}, we find that the values of intrinsic dispersion and  coefficients of the standard and Type II relations  obtained in this paper are compatible with what were determined from the combination of Pantheon SN Ia, GRBs, quasars and BAO in the $\Lambda$CDM model. While, the values of $H_0$ in the standard and Type II relations are slightly smaller than what were obtained in \citep{Dainotti2023c}. The values of $\delta$ achieved in \citep{Dainotti2023c,Lenart2023,Wang2022} and this paper are significant larger than $\delta=0.007\pm0.004$ from the golden sample of 983 quasars~\citep{Dainotti2023}. Interestingly, the constraint on $\Omega_{\mathrm{m0}}$ from quasars with  Type I relation is well consistent with  $\Omega_{\mathrm{m0}}=0.240\pm 0.064$ from 1253 gold quasar sample~\citep{Dainotti2024} and $\Omega_{\mathrm{m0}}=0.295^{+0.013}_{-0.012}$ from SN Ia + quasars~\citep{Bargiacchi2022}. Recently,  \cite{Bargiacchi2023b} found that the high-redshift evolution of our universe from GRBs and quasars in the case of   Type II relation with the effect of redshift evolution being fixed by using the EP method has a strong tension with the prediction of the $\Lambda$CDM model. Whether this tension still exists for quasars with  Type I relation is an interesting topic and needs to be investigated in the future.

\section{Conclusions}\label{sec5}

In this paper, we compare the constraints on the cosmological models from quasars with three different $L_{X}$-$L_{UV}$ relations: the standard relation proposed by \cite{Risaliti2015}, and the Type I and Type II redshift-evolutionary relations constructed by \cite{Wang2022} and~\cite{Dainotti2022}, respectively.  We employ the GP method to calibrate these relations from  the latest Pantheon+ SN~Ia data within the low-redshift region ($z<1.4$).  These results align with that reported by \cite{Wang2022}, and suggest a potential trend of redshift evolution in the $L_{X}$-$L_{UV}$ relation.

Extrapolating the calibrated relations from the low-redshift quasars to the quasars that lie in high-redshift region, we obtain the luminosity distances of quasars and then constraints from them on the spatially flat $\Lambda$CDM and $w$CDM models.
We find that the standard relation provides only a lower bound  on $\Omega_{\mathrm{m0}}$ in both the $\Lambda$CDM and  $w$CDM models. 
For the $\Lambda$CDM model, Type II relation gives effective constraints on the cosmological model parameters only when the full-redshift quasars are used.
In contrast,  Type I relation always can offer tight constraints on the cosmological model parameters, since the mean values of $\Omega_{\mathrm{m0}}$ with 1$\sigma$ CL are $\Omega_{\mathrm{m0}}=0.49^{+0.21}_{-0.33}$ for the high-redshift quasars and $\Omega_{\mathrm{m0}}=0.253^{+0.046}_{-0.067}$ for the full-redshift data.
For the $w$CDM model,  Type I relation yields the $\Omega_{\mathrm{m0}}=0.249\pm0.082$ and $w=-0.98^{+0.25}_{-0.34}$ when the full-redshift data is used, which are consistent with those obtained from the CMB data~\citep{Planck2020}.
Additionally, Type I relation always tends to give smaller values of $-2\ln\mathcal{L}$ than those of  the standard and Type II relations in both the $\Lambda$CDM and $w$CDM models.

When the additional OHD sample is considered,  the observations can  give tight constraints on the $\Lambda$CDM and $w$CDM models in all three relations, while the $\Lambda$CDM model constrained from quasars with the Type I relation still yields the tightest results. 
For this preferred model, we find that
the constraints on $\Omega_{\mathrm{m0}}$ are $0.332^{+0.050}_{-0.073}$ for the high-redshift quasars and $0.289^{+0.038}_{-0.051}$ for the full-redshift quasars,  which are consistent with the CMB results~\citep{Planck2020}.
Moreover, since the addition of the OHD sample can break the degeneracy between parameters $M$ and $H_0$, we obtain the following constraints on them.
In the $\Lambda$CDM model, the OHD measurements plus the full-redshift quasars with the Type I relation give the absolute magnitude $M$ to be $-19.321^{+0.085}_{-0.076}$, which aligns well with that obtained from SH0ES ($M=-19.253\pm{0.027}$)~\citep{Riess2022}, and  the Hubble constant $H_0$  to be $70.80\pm3.6~\mathrm{km~s^{-1}Mpc^{-1}}$, which lies between the measurements of SH0ES \citep{Riess2022}  and CMB \citep{Planck2020}.

\section*{ACKNOWLEDGMENTS}
This project was supported by  the NSFC under Grants Nos. 12275080 and 12075084,  the innovative research group of Hunan Province under Grant No. 2024JJ1006, and  the Guizhou Provincail Science and Technology Foundation (QKHJC-ZK[2021] Key 020).

\section*{DATA AVAILABILITY}
Data are available at the following references:
the 2421 X-ray and UV flux measurements of quasars
from \cite{Lusso2020},
the Pantheon+ SNe Ia sample from \cite{Scolnic2022},
and the latest OHD obtained with the cosmic chronometer method from \cite{Moresco2020}.

.



%\bibliographystyle{mnras}


%%%%%%%%%%%%%%%%%%%%%%%%%%%%%%%%%%%%%%%%%%%%%%%%%%

%%%%%%%%%%%%%%%%% APPENDICES %%%%%%%%%%%%%%%%%%%%%

%%%%%%%%%%%%%%%%%%%%%%%%%%%%%%%%%%%%%%%%%%%%%%%%%%

% Don't change these lines
\bsp	% typesetting comment
\label{lastpage}
\end{document}